\documentclass[aps,prl,showpacs,twocolumn,superscriptaddress]{revtex4}
\usepackage[toc,page]{appendix}
\usepackage{amsmath,amsthm,amssymb,mathrsfs}
\usepackage{siunitx}
\usepackage{graphicx,color,xcolor}
\usepackage{graphicx,caption,floatrow,subfig}
\usepackage[]{subfig}
\usepackage{epsfig}
\usepackage{chngcntr}
\newcommand{\beq}{\begin{equation}}
\newcommand{\eeq}{\end{equation}}
\newcommand{\bea}{\begin{eqnarray}}
\newcommand{\eea}{\end{eqnarray}}

\begin{document}

\title{ Prediction of novel `magic' angles and correlations for twisted bilayer graphene in a perpendicular electric field}

\author{Steven C. Carter}
\affiliation{School of Physics, Georgia Institute of Technology, Atlanta, GA 30332-0430, USA}
\author{Hridis K. Pal}
\affiliation{Department of Physics, University of Houston, Houston, Texas 77204, USA}
\author{M. Kindermann}
\email{markus.kindermann@physics.gatech.edu}
\affiliation{School of Physics, Georgia Institute of Technology, Atlanta, GA 30332-0430, USA}
\date{\today}

\begin{abstract}

At certain angles of rotation called `magic angles' twisted bilayer graphene features almost flat bands. The resulting strong correlations drive the system to novel phases which have been observed in experiments recently. A complete understanding of the `magic' angle physics---both at the single-particle as well as the many-particle level---is still missing, and the search is ongoing. Here, we identify a new set of `magic' angles, where locally flat bands with a variety of possible many-body instabilities arise, but where the single-particle problem admits an exact solution. This occurs in the presence of an external perpendicular electric field at multicritical Lifshitz points. At these angles, which can be quantified exactly, the central band features a monkey saddle, resulting in strong electronic correlations. 
\end{abstract}

\maketitle

When two graphene layers are rotated with respect to each other, beautiful moir{\'e} patterns appear. Such systems, commonly called twisted bilayer graphene (TBG),  host fascinating electronic properties both at the single-particle as well as at the many-particle level. They manifest nontrivial commensuration physics  \cite{mele:prb10}, topological states \cite{kindermann:prl15}, and quantum fractals \cite{kim:pnas17}, to name a few. A particularly intriguing phenomenon occurs at twist angles smaller than about $1^\circ$, when the interlayer coupling becomes nonperturbative \cite{bistritzer:pna11}: almost flat bands emerge at a set of angles termed `magic angles'  \ \cite{bistritzer:pna11}. The quenched kinetic energy at these angles results in novel, correlated phases. Recent experiments have found both Mott-like and superconducting phases in such systems \cite{cao:nat181,cao:nat182}, which has spurred frenetic activity in this field \cite{ramires2018electrically,PhysRevB.97.235453,PhysRevB.98.075154,PhysRevB.98.085436,PhysRevB.98.035425,PhysRevB.98.081102,PhysRevB.98.075109,PhysRevB.98.085435,PhysRevB.98.081410,PhysRevB.98.085144,PhysRevMaterials.2.034004,po2018origin,roy2018unconventional,baskaran2018theory,padhi2018wigner,huang2018antiferromagnetically,zhang2018low,ray2018wannier,liu2018chiral,PhysRevB.98.121406,peltonen2018mean,qiao2018heterostrain,kang2018symmetry,rademaker2018charge,kennes2018strong,isobe2018superconductivity,PhysRevX.8.031087,you2018superconductivity,pizarro2018nature,wu2018theory,pal2018magic,guinea2018electrostatic,gonzalez2018kohn,su2018spontaneous,lian2018twisted,sherkunov2018novel,sboychakov2018many,chittari2018pressure,hejazi2018multiple,lin2018minimum,laksono2018singlet,tarnopolsky2018origin,ahn2018failure,yankowitz2018tuning,venderbos2018correlations,chen2018magnetic,stauber2018linear,tang2018spin,angeli2018emergent}. Despite much progress, a complete understanding of the origin of the flat bands at  `magic' angles  is still missing. This has impeded a systematic study of correlations as well.


In this Letter we show that a new set of `magic' angles, distinct from those reported so far, arise in TBG in the presence of a perpendicular electric field. These  angles are also defined by locally flat bands leading to strong correlations, but they admit an exact solution, facilitating exploration of those correlations. The novel `magic' angles occur at  multicritical Lifshitz points, where the central band features a monkey saddle. Such a feature leads to an instability to either s-wave superconductivity, charge-density wave, spin-density wave, or ferromagnetism, depending on details of the electron-electron interaction \cite{shtyk:prb17}. These new `magic' angles are defined purely geometrically in terms of distances on the hexagonal lattice and we find them using a number theoretic approach \cite{marmon:05}. 

Realistic fermionic systems displaying a monkey saddle point are rare---only a few have been discussed in the literature so far: Majorana excitations of SU(2)-invariant  spin liquids on the triangular lattice \cite{biswas:prb11} and   Bernal bilayer graphene in a perpendicular field. The advantage with twisted graphene bilayers is that the energy scale of the monkey saddle can be enhanced by about two orders of magnitude over that of their Bernal-stacked counterparts. This strongly enhances interactions and brings the temperature required for observing the predicted correlations \cite{shtyk:prb17} into a  range readily accessible in experiments. Moreover, because the predicted `magic' angles depend on the applied electric field, this strongly correlated physics can be observed without extraordinary control of the twist angle itself.

A graphene bilayer with relative twist angle $\Theta$ and interlayer bias energy $V$ due to the perpendicular electric field  is described by a Hamiltonian
\begin{equation} 
\begin{pmatrix}
H_{1} & H_{\bot}\\
H_{\bot}^{\dagger} & H_{2}
\end{pmatrix},
\label{eq:basic_hamiltonian}
\end{equation}
where $H_{1}= v \boldsymbol{\sigma}\cdot {\bf k}$ and $H_{2}= v \boldsymbol{\sigma}\cdot {\bf k}+V $ are Dirac Hamiltonians for each individual layer and $H_{\bot}$ is the interlayer coupling. Here, we combine Pauli matrices $\sigma_{x}$ and $\sigma_{y}$ into a vector $\boldsymbol{\sigma}\equiv\begin{pmatrix}
\sigma_{x}, & \sigma_{y}
\end{pmatrix}^{T}$ and we write the Hamiltonian only for a single valley for simplicity. The interlayer Hamiltonian then has the form



\begin{equation} 
H_{\bot}({\bf r}) = \frac{\gamma}{3} \sum_{j=0}^{2} e^{i {\bf q}_{j}\cdot{\bf r}}
\begin{pmatrix}
1 & e^{- 2\pi i j/3 }\\
e^{2\pi ij/3} & 1
\end{pmatrix} .
\label{eq:interlayer_hamiltonian}
\end{equation}
The coupling energy $\gamma$ is on the order of $300\,{\rm meV}$ \cite{neto:rmp09}. 

The vectors $ {\bf q}_{j} \equiv  {\bf K}_{\theta, j}-{\bf K}_{j}$ are differences between the K-vectors ${\bf K}$ and ${\bf K}_\theta$ in the two layers, respectively, and have length $q_j=8 \pi \sin \left(\delta \theta/2\right)/\left(3 \sqrt{3}a\right)$ with ${a =1.42 \si{\angstrom}}$.

\begin{figure} 
\begin{center}
\includegraphics[width=0.8\linewidth]{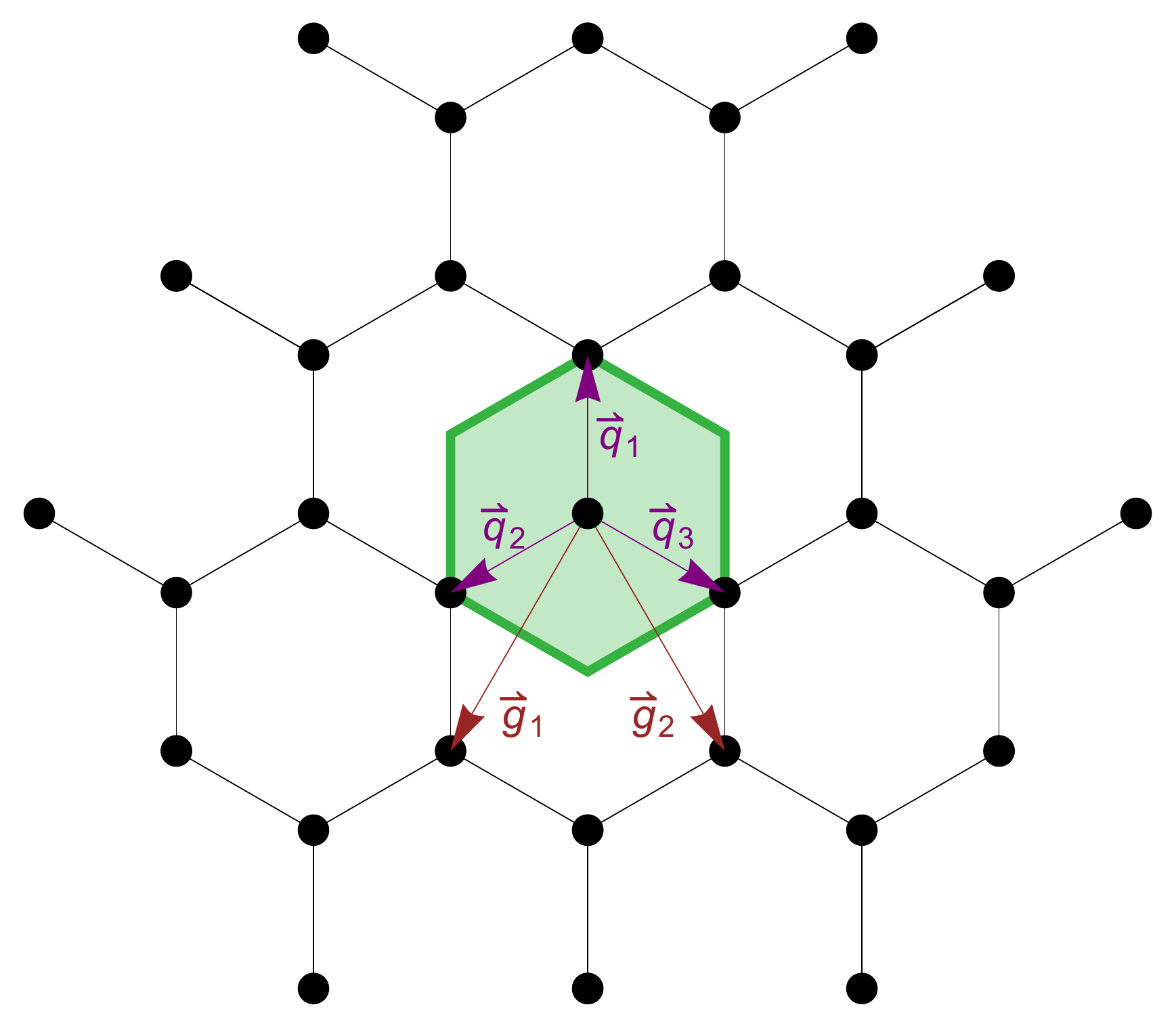}
\end{center}
\caption{(Color online) All momenta attainable from $\Gamma$ by repeated application of $H_{\bot}$. They form a hexagonal lattice in the extended zone scheme. The SBZ is indicated in green (light gray).}
\label{fig1}
\end{figure}

The interlayer Hamiltonian Eq.\ (\ref{eq:interlayer_hamiltonian}) is not translationally invariant and thus breaks momentum conservation. This makes the theory of a twisted graphene  bilayer  much harder than that of a Bernal stacked bilayer. The Hamiltonian of the twisted bilayer in our approximation, however, is periodic in real space with the period of the moir{\'e} pattern and thus does conserve crystal momentum.
In an extended Brillouin zone scheme, the momenta of states coupled to the point of zero momentum $\Gamma$ through repeated application of the interlayer Hamiltonian Eq.\  (\ref{eq:interlayer_hamiltonian}) form a hexagonal Bravais lattice with basis vectors ${\bf g}_{j}\equiv{\bf q}_{j+1}-{\bf q}_{1}$, $j\in(1,2)$. The superlattice Brillouin zone (SBZ) corresponding to the supercell defined by the moir{\'e}  pattern of the bilayer is the hexagon bounded by corners at $\pm{\bf q}_{j}$, shown in Fig. \ref{fig1}. The nonperturbative limit of a graphene bilayer without interlayer bias is reached when the interlayer coupling energy $\gamma$ becomes of order $v q_j$. Then the interlayer coupling energy $\gamma$ is able to overcome the kinetic energy cost of a change of the electron momentum by $ q_j$. A complete analytical understanding of this limit has remained elusive to date. However,  very rich phenomenology has been discovered numerically, including the mentioned `magic' angles.  \cite{bistritzer:pna11}.

\begin{figure} 
\begin{center}
\includegraphics[width=0.8\linewidth]{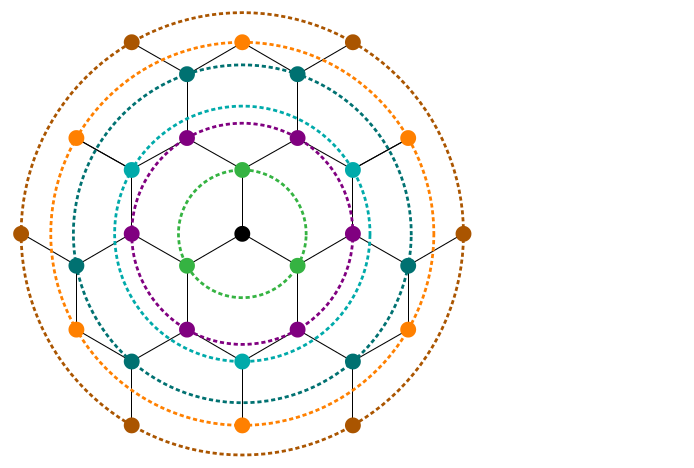}
\end{center}
\caption{(Color online) Circles indicate sets of momenta with equal magnitude on the lattice of  Fig.\ \ref{fig1}. Resonances with 3-stars -- that is circles  that cut through 3 points on that lattice -- produce a monkey saddle (cf.\ main text). }
\label{fig2}
\end{figure}

In the biased bilayer a different nonperturbative limit can be reached: consider bias energies $V_{2n+1,l} =v Q_{2n+1,l}$, where $Q_{2n+1,l}$ is the magnitude of   the momentum of a point  ${\bf Q}_{2n+1,l}$ on the lattice of Fig.\ \ref{fig2} that is connected to $\Gamma$ by a shortest path of $2n+1$ links. The index $l$ appears because in general there will be several such momentum magnitudes. Because of the $C_3$ symmetry of the lattice, there are $3m$  lattice points of that magnitude, with integer $m$.  At such bias energies   interlayer tunneling events connecting ${\bf Q}_{2n+1,l}$ to $\Gamma$ thus couple states of equal intralayer energy and such processes are resonant.  They are nonperturbative even when the bias energy $V= V_{2n+1,l}+\delta V$ deviates by $\delta V$ from its resonant value, as long as $\delta V \lesssim \gamma (\gamma/vq)^{2n}$. We take this limit, while keeping at the same time $\gamma \ll v q$, such that all other interlayer processes (corresponding to  points on the lattice Fig.\ \ref{fig2} with   distance to $\Gamma$ different from ${Q}_{2n+1,l}$) are perturbative. 
 We remark that in this limit, $\delta V \lesssim  \gamma (\gamma/vq)^{2n} \leq \gamma$ and $\gamma \ll vq$, the gaps in AB/BA-stacked regions are not able to localize the  wavefunctions  into a chiral network. This makes this limit fundamentally different from the one taken in Ref.\ \cite{sanjose:prb13}.
 
 In our limit  $\delta V \lesssim   \gamma \ll v q$ and for momenta near the $\Gamma$-point of the SBZ, the only low-energy degrees of freedom are the two sublattice amplitudes of an electron on layer $1$ with momentum near $\Gamma$ and $3m$ amplitudes for an electron to have momentum and intralayer energy near the resonant values ${\bf Q}_{2n+1,l}$  and $0$, respectively, on layer $2$. We obtain a low-energy theory of the twisted bilayer by a Schrieffer-Wolff transformation of the Hamiltonian Eq.\  (\ref{eq:basic_hamiltonian}) with low-energy space defined by those $3m+2$ amplitudes. In our limit  $ \gamma \ll v q$ we may moreover expand this effective Hamiltonian in the small parameter $\zeta=  \gamma /v q$ and we truncate this expansion at the lowest non-trivial order in $\zeta$. For the first resonance $V_{1,1}$ we find in this way a low-energy Hamiltonian
 \begin{widetext}
\begin{equation} \label{eq:super_SW}
\hspace{-20mm}H_{\text{eff}} =  \begin{pmatrix}
0 & (v k_{x}-i v k_{y}) & e^{-i \frac{2\pi}{3}}\Omega & e^{i \frac{2\pi}{3}}\Omega & \Omega \\
(v k_{x}+i v k_{y}) & 0 & \Omega & \Omega & \Omega \\
e^{i \frac{2\pi}{3}}\Omega & \Omega & v\boldsymbol{ k}\cdot \hat{q}_{1}+\delta V & 0 & 0 \\
e^{-i \frac{2\pi}{3}}\Omega & \Omega & 0 &v \boldsymbol{ k}\cdot \hat{q}_{2}+\delta V & 0 \\
\Omega & \Omega & 0 & 0 & v\boldsymbol{ k}\cdot \hat{q}_{3}+\delta V \\
 \end{pmatrix},
\end{equation} 
\end{widetext}
where $\Omega   =\gamma \sqrt{1+\frac{\sqrt{3}}{2}}$ and $\hat{q}_j$ denotes a unit vector in direction of ${\bf q}_j$.
Effective low-energy theories for other resonant spaces are found in the same way and have Hamiltonians of dimension $3m+2$. In particular, many momenta ${\bf Q}_{2n+1,l}$ with $n>0$ also have  $m=1$. We will refer to them below as ``$3$-stars.'' After a gradient expansion, their low-energy Hamiltonians  have the same form as Eq.\ (\ref{eq:super_SW}), only with renormalized values of all constants, including the resonant bias $V_{2n+1,l}$ and the $\hat{q}_j$, due to virtual interlayer processes.

While nonperturbative, as a five-dimensional matrix, $H_{\rm eff}$ has indeed exact analytic solutions in terms of special functions. Below we explore the generic properties of those solutions without referring to their explicit form, which is involved and not illuminating.

Intriguingly, our low-energy theory Eq.\ (\ref{eq:super_SW}) predicts a set of `magic' angles with flattened band for biased bilayers just as predicted in Ref.\  \cite{bistritzer:pna11} for unbiased ones. Those are the angles where perfect resonance, that is $\delta V=0$, is achieved in Eq.\ (\ref{eq:super_SW}). Electrons at the band center then have zero velocity  and  infinite mass.  This sort of localization is protected by the symmetries of the system:  In   addition to its $C_3$ symmetry, $H_{\text{eff}}$ at resonance  also has a chiral symmetry involving space inversion:
\begin{align}
\mathcal{PS}: {\bf k} \mapsto -{\bf k},\;
\psi \mapsto S \psi,
\end{align}
where
\begin{equation}
S=\begin{pmatrix}
-1 & 0 & 0 & 0 & 0\\
0 & -1 & 0 & 0 & 0\\
0 & 0 & 1 & 0 & 0\\
0 & 0 & 0 & 1 & 0\\
0 & 0 & 0 & 0 & 1
\end{pmatrix}
\end{equation}
\index{$S$}
changes the sign of energies, 
\begin{equation}
S^{\dagger}H_{\text{eff}}\left(-{\bf k}\right)S = - H_{\text{eff}}\left({\bf k}\right).
\label{eq:SP_action}
\end{equation}
Moreover, $H_{\text{eff}}$ has a low-energy band which, near $k=0$, is isolated from all other bands. This is implied by its    $k=0$,  $ \delta V=0$ eigenvalues that separate into a singlet at energy $\varepsilon=0$ and    two doublets at $\varepsilon = \pm \sqrt{3} \Omega \gamma$.  Because of its   $C_3$ symmetry, that low-energy band is bound to have zero velocity at $k=0$. Moreover, the chiral/inversion symmetry ${\cal PS}$, which precludes even powers in an expansion of the dispersion relation around $k=0$,  implies infinite mass of the charge carriers. 
The lowest term in an expansion of the dispersion relation of $H_{\rm eff}$, Eq.\ (\ref{eq:super_SW}),  at $\delta V=0$ is of order three in $k$ and takes the form of a monkey saddle:
\beq
\varepsilon=\frac{v^3}{12\Omega^2}\left(k_x^3 - 3 \sqrt{3} k_x^2 k_y - 3 k_x k_y^2 + \sqrt{3} k_y^3\right).
\eeq

Our exact solution to the low-energy physics of this system in the regime we considered allows us to not only  predict the existence of these `magic' angles, but also give an exhaustive list of their exact values. First, we recall that Eq.\ (\ref{eq:super_SW})  describes the low-energy physics at resonance with all 3-stars. Therefore, a flattened band  ftarises at all angles that have perfect resonance with a 3-star. 
On the other hand, one can see readily that isolated, flattened bands do not exist at resonance with stars that have $m>1$:  The  effective Hamiltonians for the corresponding resonant spaces at $k=0$ have rank 4, but dimension $3m+2>5$.  Therefore they have several low-energy bands that are degenerate at $k=0$ and the above reasoning does not apply. Hence, to find all `magic' angles with the above flattened bands, one needs to find the radii of all 3-stars  on the hexagonal lattice.  This can be done using a number theoretic approach outlined in Ref.\ \cite{marmon:05}: one represents points on the hopping lattice Fig.\ \ref{fig2} by the numbers corresponding to those points in the complex plane. The answer to our question is then found by prime factorization in the ring of those numbers.

We find in this way that a twisted bilayer has infinitely many  resonances with $m=1$ that occur at  values of the bias voltage
\begin{equation}
\overset{*}{V} = \hbar v_{F} \left\lvert{\bf q}_{1}\right\rvert\prod\limits_{j} m_{j}^{\mu_{j}},
\label{eq:magic_bias}
\end{equation}
where $\mu_{j}$ are arbitrary positive integers and the $m_{j}$ are primes $\equiv2\pmod{3}$. See Supplemental Material at [] for a derivation of Eq.\ (\ref{eq:magic_bias}) using the results of Ref.\ \cite{marmon:05}. At a given bias voltage $V$ there are accordingly infinitely many `magic' angles
\begin{equation}
\overset{*}{\delta \Theta} =2 \arcsin \frac{3\sqrt{3} V a}{4 h v_{F} \prod\limits_{j} m_{j}^{\mu_{j}}}
\label{eq:magic_angle}
\end{equation}
featuring a monkey saddle.

\begin{figure}
\begin{center}
\includegraphics[width=1\textwidth]{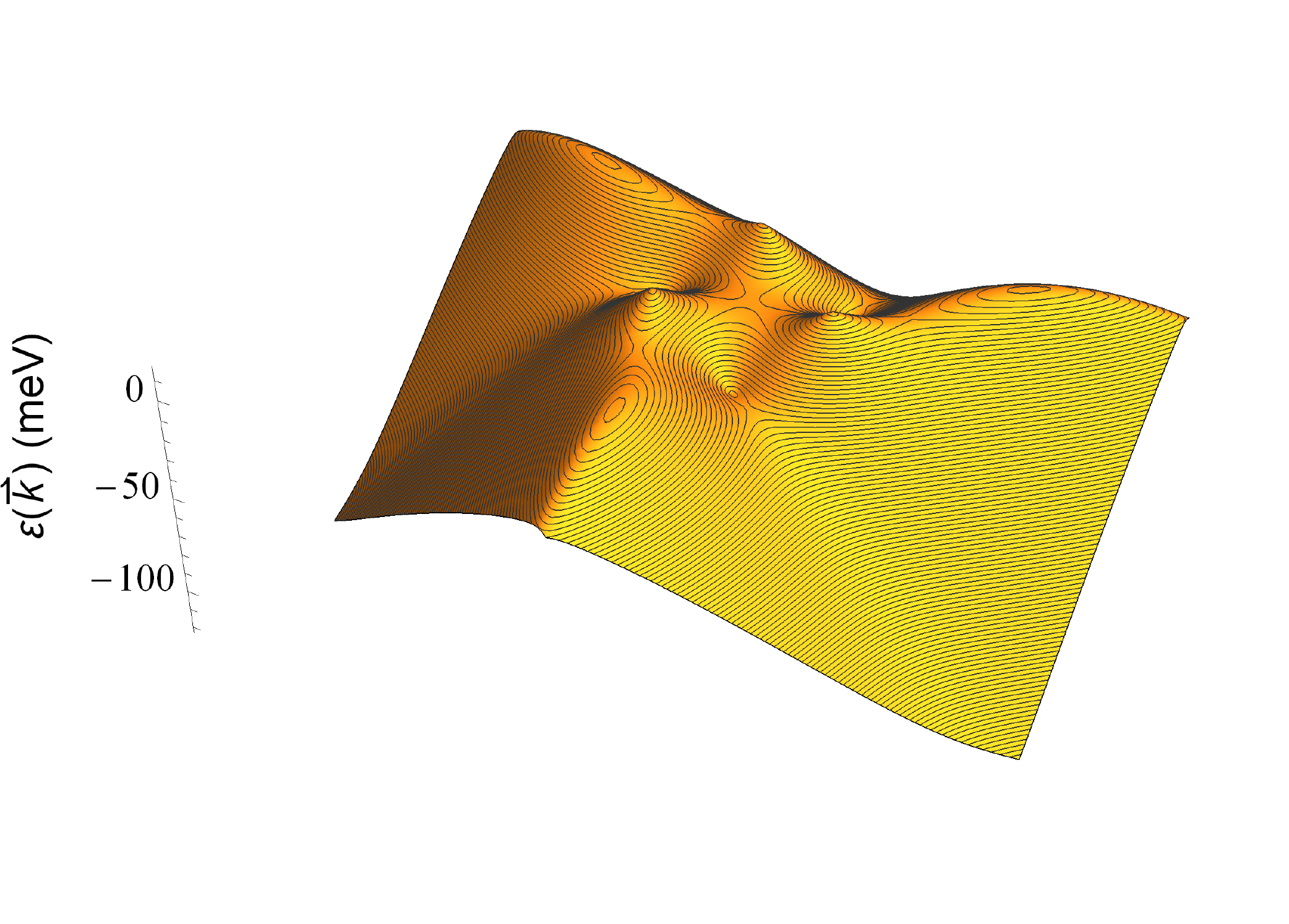}
\end{center}
\caption{(Color online) Low-energy band of   a twisted graphene bilayer with perpendicular electric field. The spectrum  is not obtained in the low-energy approximation Eq.\  (\ref{eq:super_SW}), but with a converged truncation of Eq.\  (\ref{eq:basic_hamiltonian}) of dimension $170$.  A ``monkey-saddle'' with zero gradient and zero curvature at ${\bf k}=0$  is evident.   Parameters:  $\theta=1.02 ^\circ$, $\gamma=20 \si{meV}$, $\delta V = 0$, $V=200 \si{meV}$. The domain is the full SBZ, the meshlines   are level sets of energy. }
 \label{fig3}
\end{figure}

Fig.\ \ref{fig3} depicts the lowest energy band of the system at resonance with the first 3-star. This figure is generated with a converged truncation of the full Hamiltonian Eq.\  (\ref{eq:basic_hamiltonian}) rather than the effective theory Eq.\  (\ref{eq:super_SW}).  Its  ``monkey saddle''  with zero gradient and curvature at ${\bf k}=0$ is evident. Closer inspection shows that at resonance the system is at a multicritical Lifshitz point: the crossing of two  transition lines in the space of bias voltage versus chemical potential (each line identifying a Lifshitz transition  between Fermi surfaces topologically equivalent to a single circle and to three circles, respectively). This is the same mechanism by which monkey saddles appear also in the dispersion of  Bernal bilayers with a perpendicular field \cite{shtyk:prb17}. Differently from that case, however, here the singularity in the density of states is in addition protected by the symmetries our effective theory, as discussed above.

The system has a second monkey saddle in the valley not considered explicitly here, with inverted dispersion relation.  Such pair of non-nested monkey saddles has been predicted in Ref.\   \cite{shtyk:prb17} to have a many-body instability to either  s-wave superconductivity, charge-density wave, spin-density wave, or ferromagnetism, depending on the space-dependence of the electron-electron interaction.  At nonzero $\delta V$ the charge carriers become mobile with effective mass $m^*=3(2+\sqrt{3})\gamma^2/v^2\delta V$ at the band center, tunable by the bias voltage $\delta V$. This allows great experimental control over the effective interaction strength and systematic exploration of its impact on correlations in the system.

The regime $\delta V \lesssim \gamma \ll v q$ requires bias voltages $V \approx vq \gg \gamma$ and cannot be reached experimentally in a graphene bilayer at small rotation angle, where $\gamma \approx 300 \,{\rm meV}$ \cite{neto:rmp09}. We propose two ways to reach this limit. The first one is in a heterostructure that has an h-BN layer sandwiched between two graphene layers \cite{britnell:nal12}. Since h-BN is an insulator, such a configuration at low-energies effectively acts as a graphene bilayer with reduced interlayer coupling.  Density functional theory calculations of the bandstructure of such systems \cite{amithray:jap16} predict an effective coupling of order $\gamma \approx 20\, {\rm meV}$ between the graphene layers \footnote{This agrees with the order of magnitude of both the expected direct graphene-to-graphene coupling by wavefunction overlap between carbon atoms in the two layers \cite{dresselhaus:aip02,nilsson:prb08,brandt:bo88,chung:jms02} and the mediated coupling by virtual hopping onto a B or a N atom  $\gamma' \approx \gamma_{\rm g h-BN}^2/\varepsilon_{\rm B} + \gamma_{\rm g h-BN}^2/\varepsilon_{\rm N} $ (here, $ \gamma_{\rm g h-BN}\approx .25 \, {\rm eV}$ is the coupling between a graphene layer and the h-BN layer and $\varepsilon_{\rm B}\approx 3.16\,{\rm eV}$, $\varepsilon_{\rm N}\approx -1.50\,{\rm eV}$ are the energies of the relevant orbitals in the B and N atoms \cite{slawinska:prb102})}. Therefore, our limit in such sandwiches is  reached at experimentally attainable bias voltages  $V \gtrsim 100 \, {\rm meV}$. Insertion of additional h-BN layers further reduces the required bias fields. The second way to achieve the limit  
$\delta V \lesssim \gamma \ll v q$ is by choosing a rotation angle near $\theta = 38.21^\circ$. In that case the dominant interlayer coupling is of higher order and the above results apply with reduced interlayer coupling $\gamma\approx 7\;{\rm meV}$ and $q=\sqrt{3}\delta K$ \cite{pal:14}.

The monkey saddle of Fig.\ \ref{fig3}  has the cubic form assumed in the renormalization group analysis of Ref.\ \cite{shtyk:prb17} up to energy scale $\gamma$.  In the proposed twisted sandwich of a single  h-BN layer between two graphene layers scaling can thus be observed up to energies of order $20 \, {\rm meV}$, which is  about two more orders of magnitude than in Bernal graphene bilayers \cite{shtyk:prb17}. This strongly enhances many-body interactions and brings the predictions of Ref.\ \cite{shtyk:prb17} into immediate experimental reach.

In conclusion, we have identified a new set of `magic' angles in twisted graphene bilayers in the presence of a perpendicular electric field. Unlike in the unbiased case, the system admits an exact solution at these  angles. The band structure becomes locally flat with a pair of non-nested ``monkey saddles'' and  many-body instabilities as discussed in Ref.~\cite{shtyk:prb17}. The perpendicular field gives great experimental control and allows  observation of this physics across a wide range of twist angles by tuning the  field to  multicritical Lifshitz points. The resulting correlations can be observed at  temperatures about two orders of magnitude higher than in previously proposed realizations  \cite{shtyk:prb17}.

\end{document}


\bibliographystyle{apsrev}

 \section{ Supplemental Material A}

Here we derive the `magic' bias voltages given in Eq.\ (8) of the main text. We heavily rely on a proof given in Ref.\   \cite{marmon:05}. To make contact with that paper we define two lattices $ \mathbb{E}_j$, $ j\in\{0,1\}$,  spanned by normalized vectors $\vec{\phi}_j=\vec{q}_j/|\vec{g}_j|$ and  $\vec{\Phi}_j=\vec{g}_j/|\vec{g}_j|$,
\begin{align}
 \mathbb{E}_{j} =\{n\vec{\Phi}_{1}+m\vec{\Phi}_{2}+j\vec{\phi}_{1} | n,m\in\mathbb{Z}\},
\label{eq:E0_E1_sets}
\end{align}
where $\vec{q}_j $ and $\vec{g}_j$ are defined in the main text and depicted in Fig.\ 1. 
We furthermore introduce a norm
\begin{equation}
N(\vec{\eta})=\vec{\eta}\cdot\vec{\eta}
\end{equation}
 on $ \mathbb{E}_j$ and observe that  
\begin{equation} \label{phi}
\vec{\phi}_{1} = -\left(\vec{\Phi}_{1}+\vec{\Phi}_{2}\right)/3.
\end{equation}

\begin{thm}\label{thm:natural}
 For every $\vec{\eta} \in \mathbb{E}_1$ one has $3N(\vec{\eta})  \equiv 1 \pmod{3}$.

\begin{proof}
Let $\vec{\eta} \in \mathbb{E}_1$. Then
\begin{align*}
3 N( \vec{\eta}) &=3 \left(n \vec{\Phi}_{1} + m \vec{\Phi}_{2} +  \vec{\phi}_{1}\right)\cdot \left(n \vec{\Phi}_{1} + m \vec{\Phi}_{2} +  \vec{\phi}_{1}\right)=
3\left(n^{2} + m^{2}  + n m -  n - m\right)+1.\\
\end{align*}
\end{proof}
\end{thm}

\begin{thm}\label{thm:triple}
For every $\vec{\eta} \in \mathbb{E}_j$ and $p \in \mathbb{Z}$, $3p \vec{\eta} \in \mathbb{E}_{0}$.
\begin{proof}
For $\vec{\eta} \in \mathbb{E}_j$ have
\begin{align*}
   3p \vec{\eta} &= p \left[3n \vec{\Phi}_{1} + 3m \vec{\Phi}_{2} + j \left(3\vec{\phi}_{1}\right)\right]= p \left[3n \vec{\Phi}_{1} + 3m \vec{\Phi}_{2} - j \left(\vec{\Phi}_{1}+\vec{\Phi}_{2}\right)\right] \in \mathbb{E}_{0}
\end{align*}
\end{proof}
\end{thm}

\begin{thm} \label{thm:magic_lengths}
There are exactly 3 distinct solutions $\vec{\lambda} \in \mathbb{E}_{1}$ to $\sqrt{N(\lambda)}= L$ if and only if $L=3^{-1/2}\Pi \,m_{j}^{\mu_{j}}$ with $ \mu_{j}\in\mathbb{N}$ and primes $m_{j}\equiv2\pmod{3}$. 

\begin{proof}
Let there be exactly
3 vectors $\vec{\lambda}  \in \mathbb{E}_{1}$ that have length $\sqrt{N(\lambda )}= L$.
Define $\vec{\Lambda}  \equiv 3\vec{\lambda}$. By Theorem \ref{thm:triple}, $\vec{\Lambda}  \in \mathbb{E}_{0}$.
Because of the inversion symmetry of  $\mathbb{E}_{0}$, also $-\vec{\Lambda} \in \mathbb{E}_{0}$. Thus, to every 3-star in $\mathbb{E}_{1}$ of radius $L$ we have at least 6 vectors in $\mathbb{E}_{0}$ that have length $3L$. We first show that they form a 6-star, that is we show that there are no more than $6$ vectors in $\mathbb{E}_{0}$ with length $3L$.

 Consider an arbitrary vector  $\vec{\Lambda}^{\prime} = l \vec{\Phi}_{1} + (l+d) \vec{\Phi}_{2} \in \mathbb{E}_{0}$ with length $3L$, that is $N(\vec{\Lambda}^{\prime})= 9L^2=9N(\vec{\lambda})$. We have then
\begin{align*}
3 N(\vec{\lambda})&=
N(\vec{\Lambda}^{\prime})/3=   \left(
l^{2} \left\lvert\vec{\Phi}_{1}\right\rvert^{2} + (l+d)^{2} \left\lvert\vec{\Phi}_{2}\right\rvert^{2}  + 2 l(l+d) \vec{\Phi}_{1}\cdot \vec{\Phi}_{2}  \right)/3 =   l^{2}+ld+d^2/3.\\
\end{align*}
By Theorem \ref{thm:natural},  $3N(\vec{\lambda})  \in \mathbb{N}_0$. It follows that $d=3r$ with integer $r$ and thus $3 N(\vec{\lambda})= l^{2}+3rl+3 r^2$. Furthermore, by the same theorem, $3N(\vec{\lambda}) \equiv 1 \pmod{3}$. We conclude that $l \neq 0 \pmod{3}$ and we can write   $l=3s-t$ with    $t\in \{-1,1\}$ and $s \in \mathbb{Z}$. Now consider the vector $\vec{\lambda}^{\prime}=\vec{\Lambda}^{\prime}/3$:
\begin{align*}
\vec{\lambda}^{\prime}
&=\left[ l \vec{\Phi}_{1}+ \left(l+d\right) \vec{\Phi}_{2}\right]/3=\left[ (3s-t) \left( \vec{\Phi}_{1}+\vec{\Phi}_{2}\right)+3r \vec{\Phi}_{2}\right]/3= s \left( \vec{\Phi}_{1}+\vec{\Phi}_{2}\right)+  t  \vec{\phi}_{1}+r \vec{\Phi}_{2},
\end{align*}
where the last equality follows from Eq.\ (\ref{phi}). We conclude that either $\vec{\lambda}^{\prime} \in \mathbb{E}_{1}$ (if $t=1$), or $-\vec{\lambda}^{\prime} \in \mathbb{E}_{1}$ (if $t=-1$). We see that indeed all vectors $\vec{\Lambda}^{\prime} \in \mathbb{E}_{0}$ with length $3L$ derive from a vector $\vec{\lambda} \in \mathbb{E}_{1}$ of length $L$ by multiplication with $3$ or $-3$. There are only $3$ such vectors $\vec{\lambda} \in \mathbb{E}_{1}$ by our assumption. Therefore, the vectors  $\vec{\Lambda}^{\prime} $ indeed form a $6$-star in $ \mathbb{E}_{0}$ and $\vec{\Lambda}$ is part of that $6$-star.

Using prime factorization and partitioning of prime factors into equivalence classes modulo 3, the radii $N(\vec{\Lambda})$ of all stars in $ \mathbb{E}_{0}$ are found in Ref.\ \cite{marmon:05} to equal
\begin{align} \label{eq:modulofactor}
N(\vec{\Lambda}) &= 3^{\beta}\prod\limits_{i} n_{i}^{\nu_{i}}
\prod\limits_{j} m_{j}^{\kappa_{j}} \nonumber\\
\end{align}
with unique $\beta,\nu_{k},\mu_{k} \in \mathbb{N}$ and primes $n_{k} \equiv 1 \pmod{3}$, $m_{k} \equiv 2 \pmod{3}$.\\
By theorem \ref{thm:natural},   $N(\vec{\Lambda})/3=3N(\vec{\lambda}) \equiv 1 \pmod{3}$, so $\beta=1$.
Furthermore, it is shown in Ref.\   \cite{marmon:05} that all $\kappa_j=2\mu_j$ are even numbers.

According to Eq.\ (7) of Ref.\   \cite{marmon:05}   the number of lattice sites in $\mathbb{E}_{0}$ on a circle of radius $N(\vec{\Lambda}) $ is
\begin{equation}\label{eq:circlenumber}
\mathfrak{N}\left[N(\vec{\Lambda}) \right]= 6 \cdot \prod\limits_{n_{i}} \left(\nu_{i}+1\right),
\end{equation}
with the $n_{i},\nu_{i}$ of Eq.\ \ref{eq:modulofactor}. Immediately we see that $\mathfrak{N}\left[N(\vec{\Lambda}) \right]=6$ requires all the $\nu_{i}$ to be zero. Thus, indeed,
\begin{align} \label{L}
L=\frac{1}{3} \sqrt{N(\vec{\Lambda})} = 3^{-1/2}\, \prod\limits_{j} m_{j}^{\mu_{j}}
\end{align}
with   $\mu_{k} \in \mathbb{N}$ and primes  $m_{k} \equiv 2 \pmod{3}$,  as claimed.

It remains to be shown that each $6$-star in $ \mathbb{E}_{0}$ with a radius  $3L $  given by Eq.\ (\ref{L}) also corresponds to a $3$-star in $ \mathbb{E}_{1}$.  To show this we use that according to Ref.\ \cite{marmon:05}  circles containing exactly 6 solutions in $\mathbb{E}_{0}$ must include a solution $\vec{\Lambda}_{y}\in \mathbb{E}_{0}$ such that $\vec{\Lambda}_{y}\cdot \hat{x}=0$, that is $\vec{\Lambda}_{y} = u \left(\vec{\Phi}_{1} + \vec{\Phi}_{2}\right) $ with integer $u$. Moreover, $L=\sqrt{N(\vec{\Lambda}_{y} )}/3=3^{-1/2}u$ and comparison with Eq.\ (\ref{L}) shows that $u$ cannot be divisible by $3$. Using Eq.\ (\ref{phi}) we have  $\vec{\Lambda}_{y} = 3u\vec{\phi}_{1} $ and thus it follows as above that either $\vec{\Lambda}_{y} /3$ or $-\vec{\Lambda}_{y}/3 $ are in  $ \mathbb{E}_{1}$. By the $C_3$ symmetry of the lattices  $ \mathbb{E}_{j}$ there are at least two more vectors of the same length. By   Theorem \ref{thm:triple} there cannot be more than three such vectors, since $\vec{\Lambda}_{y} $ is part of a $6$-star. Thus either $\vec{\Lambda}_{y} /3$ or $-\vec{\Lambda}_{y}/3 $ are indeed part of a $3$-star of radius $L$ in $ \mathbb{E}_{1}$.

\end{proof}
\end{thm}

Theorem \ref{thm:magic_lengths} together with the resonance condition $V_{2n+1,l} =v Q_{2n+1,l}$ immediately results in Eq.\ (8) of the main text.

\bibliographystyle{apsrev}  
\bibliography{/Users/mkinderman3/Documents/tex/refs/refs_graphene,/Users/mkinderman3/Documents/tex/refs/refs_kindermann}